\documentclass[12pt]{article}
\usepackage{amssymb,epsfig}

\begin{document}

\hspace*{\fill}TGU-30

\hspace*{\fill}UWThPh-2003-11

\hspace*{\fill}ZU-TH 09/03

\hspace*{\fill}LC-TH-2003-041

\hspace*{\fill}hep-ph/0306281

\vspace{1cm}
\begin{center}
\textbf{\Large Impact of SUSY CP Phases on Stop and Sbottom Decays in the MSSM}

\vspace{10mm}
{\large
A.~Bartl$^a$, S.~Hesselbach$^a$, K.~Hidaka$^b$, T.~Kernreiter$^a$,
W.~Porod$^c$
}

\vspace{5mm}
\itshape
$^a$Institut f\"ur Theoretische Physik, Universit\"at Wien, A-1090
  Vienna, Austria\\
$^b$Department of Physics, Tokyo Gakugei University, Koganei, Tokyo
  184-8501, Japan\\
$^c$Institut f\"ur Theoretische Physik, Universit\"at Z\"urich,
  CH-8057 Z\"urich, Switzerland

\end{center}

\vspace{5mm}
\begin{abstract}
We study the decays of top squarks and bottom squarks in the Minimal
Supersymmetric Standard Model with complex parameters $A_t$, $A_b$ and
$\mu$. In a large region of the supersymmetric parameter space the
branching ratios of $\tilde{t}_1$ and $\tilde{b}_1$ show a pronounced
phase dependence. This could have an important impact on the search
for $\tilde{t}_1$ and $\tilde{b}_1$ at a future linear collider and
on the determination of the supersymmetric parameters.
\end{abstract}

\section{Introduction}

So far most phenomenological studies on production and decay of
supersymmetric (SUSY) particles have been performed within the
Minimal Supersymmetric Standard Model \linebreak (MSSM) \cite{mssm}
with real SUSY parameters.
In this contribution we analyze the decays of $\tilde{t}_1$ and
$\tilde{b}_1$ in the MSSM with complex SUSY parameters.
The lighter squark mass eigenstates may be relatively light and could
be thoroughly studied at an $e^+e^-$ linear collider. 

In the third generation sfermion sector the mixing between the left
and right states 
cannot be neglected because of the effects of the large Yukawa
couplings. The left-right mixing terms in the squark mass matrix
depend on the higgsino mass parameter $\mu$ and the trilinear scalar
couplings $A_q$, $q=t,b$, which may be complex in general.
In mSUGRA-type models the phase $\varphi_{\mu}$ of $\mu$ turns out to
be restricted by the experimental data on electron, neutron and mercury
electric dipole moments (EDMs) to a range $|\varphi_{\mu}| \lesssim 0.1$
-- 0.2 for an universal scalar mass parameter $M_0 \lesssim 400$~GeV
\cite{edm1, edm2,Abel:2001vy}.
However, the restriction due to the electron
EDM can be circumvented if complex lepton flavour violating terms
are present in the slepton sector \cite{sleptonedm}.
The phases of the parameters $A_{t,b}$ are not restricted at
one-loop level by the EDM data but only at two-loop level, resulting
in much weaker constraints on these phases \cite{edmAf}. 

Analyses of the decays of the 3rd generation squarks
$\tilde{t}_{1,2}$ and $\tilde{b}_{1,2}$ in the MSSM with real
parameters were performed in Refs.~\cite{realdecay1,realdecay2} and
phenomenological studies of production and decay of these particles at
future $e^+e^-$ colliders in Ref.~\cite{realprod}. A detailed study how to
determine $A_t$ and $A_b$ in the real MSSM with help of the measured
polarization of final state top quarks was performed in Ref.~\cite{polprod}.
Recently the influence of complex phases on the phenomenology of third
generation sleptons has been studied in \cite{slepton}.

In this article we study the effects of the complex phases of $A_t$,
$A_b$ and $\mu$ on the partial decay widths and branching ratios of
$\tilde{t}_1$ and $\tilde{b}_1$.
We assume, that the gaugino mass parameters are real.
Especially the effects of the
possibly large phases of $A_t$ and $A_b$ can be quite strong, which
would have an important impact on the search for $\tilde{t}_1$
and $\tilde{b}_1$ at a future $e^+e^-$ linear collider.

\section{\boldmath $\tilde{q}_L$-$\tilde{q}_R$ mixing}

The left-right mixing of the stops and sbottoms is described by a
hermitian $2 \times 2$ mass matrix, which in the basis
$(\tilde{q}_L,\tilde{q}_R)$ reads
\begin{equation} \label{squarkmixmatrix}
\mathcal{L}^{\tilde{q}}_M = - (\tilde{q}_L^*,\tilde{q}_R^*)
 \left(\begin{array}{cc}
    M_{\tilde{q}_{LL}}^2 & M_{\tilde{q}_{LR}}^2\\
    M_{\tilde{q}_{RL}}^2 & M_{\tilde{q}_{RR}}^2
 \end{array}\right)
 \left(\begin{array}{c} \tilde{q}_L \\ \tilde{q}_R \end{array}\right),
\end{equation}
with
\begin{eqnarray}
M_{\tilde{q}_{LL}}^2 & = & M_{\tilde{Q}}^2
  + (T_q^3 - Q_q \sin^2 \theta_W) \cos 2\beta\; m_Z^2 + m_q^2, \\
M_{\tilde{q}_{RR}}^2 & = & M_{\tilde{Q'}}^2
  + Q_q \sin^2 \theta_W \cos 2\beta\; m_Z^2 + m_q^2, \\
M_{\tilde{q}_{RL}}^2 & = & (M_{\tilde{q}_{LR}}^2)^* =
  m_q \left(A_q - \mu^* (\tan\beta)^{-2 T_q^3}\right), \label{mLRterm}
\end{eqnarray}
where $m_q$, $Q_q$ and $T_q^3$ are the mass, electric charge and weak
isospin of the quark $q=b,t$. $\theta_W$ denotes the weak mixing angle,
$\tan\beta = v_2/v_1$ with $v_1$ ($v_2$) being the vacuum expectation
value of the Higgs field $H^0_1$ ($H^0_2$) and
$M_{\tilde{Q'}}= M_{\tilde{D}}$ ($M_{\tilde{U}}$) for $q = b$ ($t$).
$M_{\tilde{Q}}$, $M_{\tilde{D}}$, $M_{\tilde{U}}$, $A_b$ and $A_t$
are the soft SUSY-breaking parameters of the stop and sbottom system.
In case of complex parameters $\mu$ and $A_q$ the off-diagonal elements
$M_{\tilde{q}_{RL}}^2 = (M_{\tilde{q}_{LR}}^2)^*$ are also complex
with the phase
\begin{equation} \label{defphisquark}
\varphi_{\tilde{q}} = \arg\left[M_{\tilde{q}_{RL}}^2\right]
  = \arg\left[A_q - \mu^* (\tan\beta)^{-2 T_q^3}\right].
\end{equation}
$\varphi_{\tilde{q}}$ together with the squark mixing angle
$\theta_{\tilde{q}}$ fixes the mass eigenstates of the squarks
\begin{eqnarray}
\tilde{q}_1 & = & e^{i\varphi_{\tilde{q}}} \cos\theta_{\tilde{q}}\, \tilde{q}_L
    + \sin\theta_{\tilde{q}}\, \tilde{q}_R \,,\\
\tilde{q}_2 & = &  -\sin\theta_{\tilde{q}}\, \tilde{q}_L
    + e^{-i\varphi_{\tilde{q}}} \cos\theta_{\tilde{q}}\, \tilde{q}_R
\end{eqnarray}
with
\begin{equation}
\cos\theta_{\tilde{q}} =
 \frac{-|M_{\tilde{q}_{LR}}^2|}{\sqrt{|M_{\tilde{q}_{LR}}^2|^2
  + (m_{\tilde{q}_1}^2 - M_{\tilde{q}_{LL}}^2)^2}},\quad
\sin\theta_{\tilde{q}} =
 \frac{M_{\tilde{q}_{LL}}^2 - m_{\tilde{q}_1}^2}{\sqrt{|M_{\tilde{q}_{LR}}^2|^2
  + (m_{\tilde{q}_1}^2 - M_{\tilde{q}_{LL}}^2)^2}}
\end{equation}
and the mass eigenvalues
\begin{equation}
m_{\tilde{q}_{1,2}}^2 = \frac{1}{2} \left(
  (M_{\tilde{q}_{LL}}^2 + M_{\tilde{q}_{RR}}^2) \mp
  \sqrt{(M_{\tilde{q}_{LL}}^2 - M_{\tilde{q}_{RR}}^2)^2 +
    4 |M_{\tilde{q}_{LR}}^2|^2}\right).
\end{equation}

\section{Numerical results}

In this section we will present numerical results for the phase
dependences of the $\tilde{t}_1$ and $\tilde{b}_1$ partial decay
widths and branching ratios. We calculate the partial decay widths in
Born approximation. It is known that in some
cases the one-loop SUSY QCD corrections are important. The analyses of
\cite{realdecay2, mbrun, mbrun2} suggest that a significant part of the
one-loop SUSY QCD
corrections to certain $\tilde{t}_1$ and $\tilde{b}_1$ partial decay
widths can be
incorporated by using an appropriately corrected bottom quark mass. In
this spirit we calculate the tree-level widths of the $\tilde{t}_1$ and
$\tilde{b}_1$ decays by using on-shell masses for the kinematic factors,
whereas we take running masses for the top and bottom quark
for the Yukawa couplings. For definiteness we take
$m_t^\mathrm{run}(m_Z) = 150$~GeV,
$m_t^\textrm{\scriptsize on-shell} = 175$~GeV,
$m_b^\mathrm{run}(m_Z) = 3$~GeV and
$m_b^\textrm{\scriptsize on-shell} = 5$~GeV.
This approach leads to an ``improved'' Born approximation, which takes
into account an essential part
of the one-loop SUSY QCD corrections to the $\tilde{t}_1$ and $\tilde{b}_1$
partial decay widths and predicts their phase dependences more accurately
than the ``naive'' tree-level calculation.

In the numerical analysis we impose the following conditions in order
to fulfill the experimental constraints:
$m_{\tilde{\chi}^\pm_1} > 103$~GeV,
  $m_{\tilde{\chi}^0_1} > 50$~GeV, $m_{H_1} > 100$~GeV,
  $m_{\tilde{t}_1, \tilde{b}_1} > 100$~GeV,
  $m_{\tilde{t}_1, \tilde{b}_1} > m_{\tilde{\chi}^0_1}$,
$\Delta\rho(\tilde{t}-\tilde{b}) < 0.0012$ \cite{deltarho}.
We also calculate the branching ratio for $b \to s \gamma$ and
compare it with the experimentally allowed range 
$2.0 \times 10^{-4} < B(b \to s \gamma) < 4.5 \times 10^{-4}$ \cite{bsgamma}.

First we discuss the dependence of the  $\tilde{t}_1$ partial decay
widths on $\varphi_{A_t}$ and $\varphi_\mu$ in two scenarios inspired
by the Snowmass Points and Slopes scenarios SPS~1a and SPS~4
\cite{sps}. For this we take the squark masses, the squark mixing
angles, $\mu$, $\tan\beta$ and $M_2$ from \cite{sps} as
input and compute $|A_t|$ from this with help of
eqs.~(\ref{squarkmixmatrix}) -- (\ref{mLRterm}).
The relevant parameters for the determination of the partial decay
widths are summarized in Table~\ref{scentab}.
Furthermore, we also look at a scenario with a light $\tilde{t}_1$ allowing
production of this particle at a 500~GeV linear collider.
When varying the phases of $A_t$ and $\mu$ we fix three squark masses
and the absolute values of the parameters at the given values,
calculating $M_{\tilde{Q}}$, $M_{\tilde{U}}$ and $M_{\tilde{D}}$
accordingly. Then the fourth squark mass $m_{\tilde{b}_2}$
($m_{\tilde{t}_2}$) in case of stop (sbottom) decays depends on the
phases and varies around the given value.

\begin{table}[ht]
\centering
\renewcommand{\arraystretch}{1.5}
\addtolength{\tabcolsep}{2mm}
\begin{tabular}{|l||c|c|c|c|}
\hline
 & SPS~1a & SPS~4 & light $\tilde{t}_1$ & light $\tilde{b}_1$ \\ \hline\hline
$m_{\tilde{t}_1}/$GeV & 379.1 & 530.6 & 240.0 & 170.0 \\ \hline
$m_{\tilde{t}_2}/$GeV & 574.7 & 695.9 & 700.0 & $\approx 729$ \\ \hline
$m_{\tilde{b}_1}/$GeV & 491.9 & 606.9 & 400.0 & 350.0\\ \hline
$m_{\tilde{b}_2}/$GeV & $\approx 540$ & $\approx 709$ & 
  $\approx 662$ & 700.0\\ \hline
$|A_t|/$GeV & 465.5 & 498.9 & 600.0 & 600.0 \\ \hline
$|\mu|/$GeV & 352.4 & 377.0 & 400.0 & 300.0\\ \hline
$\tan\beta$ & 10 & 50 & 6 & 30\\ \hline
$M_2/$GeV & 192.7 & 233.2 & 135.0 & 200.0\\ \hline
$m_{H^\pm}/$GeV & 401.8 & 416.3 & 900.0 & 150.0\\ \hline
$\tilde{q}$ mixing & $M_{\tilde{Q}} > M_{\tilde{U}}$ &
  $M_{\tilde{Q}} > M_{\tilde{U}}$ & 
  $M_{\tilde{Q}} > M_{\tilde{U}}$ &
  $M_{\tilde{Q}} > M_{\tilde{D}}$\\ \hline
\end{tabular}
\caption{\label{scentab}Relevant parameters in scenarios used to
discuss the stop and sbottom decays.}
\end{table}

\begin{figure}[p]
\centering
\begin{picture}(16,17)
%\put(0,0){\framebox(16,17){}}

\put(-.1,0){\epsfig{file=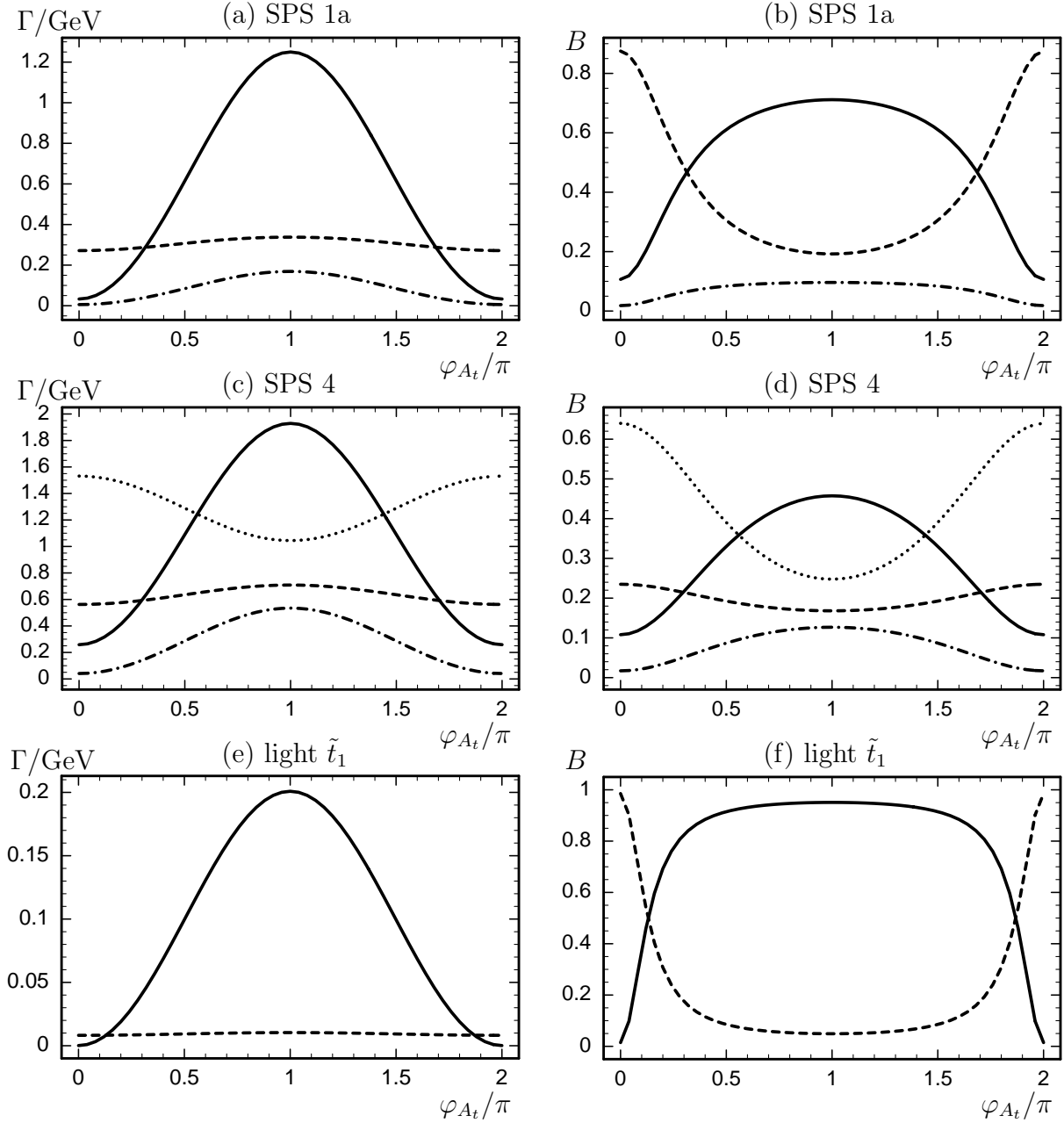}}

\end{picture}
\caption{\label{decayphiAt}(a), (c), (e) Partial decay widths and (b), (d),
  (f) branching ratios of the decays
$\tilde{t}_1 \to \tilde{\chi}^+_1 b$ (solid),
$\tilde{t}_1 \to \tilde{\chi}^0_1 t$ (dashed),
$\tilde{t}_1 \to \tilde{\chi}^+_2 b$ (dashdotted) and
$\tilde{t}_1 \to \tilde{\chi}^0_2 t$ (dotted)
in the SPS~1a and SPS~4 inspired scenarios and the scenario with a light
$\tilde{t}_1$ defined in Table~\ref{scentab} for
$\varphi_{A_b}=\varphi_{\mu}=0$.}

\end{figure}

In Fig.~\ref{decayphiAt} we show the partial decay widths and
branching ratios for $\tilde{t}_1 \to \tilde{\chi}^+_1 b$,
$\tilde{t}_1 \to \tilde{\chi}^0_1 t$,
$\tilde{t}_1 \to \tilde{\chi}^+_2 b$ and
$\tilde{t}_1 \to \tilde{\chi}^0_2 t$
as a function of $\varphi_{A_t}$ in the scenarios of
Table~\ref{scentab} for $\varphi_{A_b} = \varphi_{\mu} = 0$. 
Fig.~\ref{decayphiAt} (a) and (b) are for
the SPS~1a inspired scenario. Here the partial decay
widths of the chargino channels
$\tilde{t}_1 \to \tilde{\chi}^+_{1,2} b$
show a significant $\varphi_{A_t}$ dependence, whereas 
$\tilde{t}_1 \to \tilde{\chi}^0_1 t$ has only a weak phase dependence.
For $\varphi_{A_t}\approx 0, 2\pi$
the decay into $\tilde{\chi}^0_1 t$ dominates, whereas for
$\varphi_{A_t}\approx \pi$ the decay into $\tilde{\chi}^+_1 b$ has the
largest branching ratio.
In Fig.~\ref{decayphiAt} (c) and (d) we show the decays in
the SPS~4 inspired scenario. Here all four partial decay widths
contribute with comparable size. Again the chargino channels 
$\tilde{t}_1 \to \tilde{\chi}^+_{1,2} b$
show the largest $\varphi_{A_t}$ dependence. However, for
$\varphi_{A_t}\approx 0, 2\pi$ the decay 
$\tilde{t}_1 \to \tilde{\chi}^0_2 t$ dominates.
In the scenario with a light $\tilde{t}_1$
(Fig.~\ref{decayphiAt} (e) and (f)) only the decay channels
into $\tilde{\chi}^+_1$ and $\tilde{\chi}^0_1$ are open.
Also in this scenario $\Gamma(\tilde{t}_1 \to \tilde{\chi}^+_{1} b)$
shows this clear $\varphi_{A_t}$ dependence, resulting in
$B(\tilde{t}_1 \to \tilde{\chi}^0_1 t)$ dominating at
$\varphi_{A_t}\approx 0, 2\pi$ and $B(\tilde{t}_1 \to \tilde{\chi}^+_{1} b)$
dominating at $\varphi_{A_t}\approx \pi$.
The decay pattern, especially of $\tilde{t}_1 \to \tilde{\chi}^+_1 b$,
can be explained in the following way:
In all scenarios we have $|A_t| \gg |\mu|/\tan\beta$,
therefore $\theta_{\tilde{t}}$ depends
only weakly on $\varphi_{A_t}$. However, $\varphi_{\tilde{t}}\approx
\varphi_{A_t}$ (see eq.~(\ref{defphisquark})), which causes
the clear $1 - \cos \varphi_{A_t}$
behavior of $\Gamma(\tilde{t}_1 \to \tilde{\chi}^+_1 b)$.
We have calculated $B(b \to s\gamma)$ in all three scenarios. In the case
of SPS~1a we obtain $B(b \to s\gamma)$ in the experimentally allowed
range for $0.5\pi < \varphi_{A_t} < 1.5\pi$, whereas for
$\varphi_{A_t} \approx 0, 2\pi$ it can reach values up to
$5\times 10^{-4}$. In the case of SPS~4 and the scenario with a light
$\tilde{t}_1$ the situation is quite similar, with $B(b \to s\gamma)$
reaching values of $6.5\times 10^{-4}$ and $5.3\times 10^{-4}$,
respectively, near $\varphi_{A_t} \approx 0, 2\pi$.

\begin{figure}[ht]
\centering
\begin{picture}(16,8.5)
%\put(0,0){\framebox(16,8.5){}}

\put(-.1,0){\epsfig{file=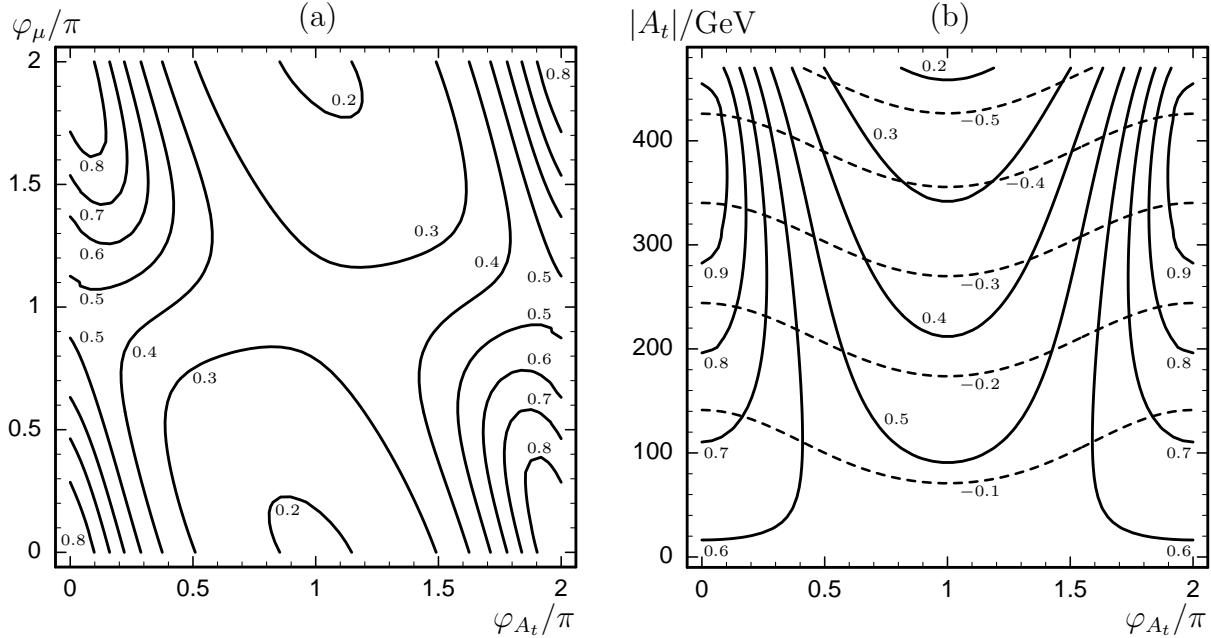}}

\end{picture}
\caption{\label{cpneut1sps1a}Contours of 
$B(\tilde{t}_1 \to \tilde{\chi}^0_1 t)$ in the SPS~1a inspired
scenario defined in Table~\ref{scentab} for 
(a) $\varphi_{A_b}=0$ and 
(b) $|A_b|=|A_t|$, $\varphi_{A_b}=\varphi_{\mu}=0$.
The dashed lines in (b) denote the contours of $\cos\theta_{\tilde{t}}$.}

\end{figure}

In Fig.~\ref{cpneut1sps1a} (a) we show a contour plot for the
branching ratio 
$B(\tilde{t}_1 \to \tilde{\chi}^0_1 t)$ as a function of $\varphi_{A_t}$
and $\varphi_\mu$ for $\varphi_{A_b}=0$ in the SPS~1a inspired scenario.
The $\varphi_{A_t}$ dependence is
stronger than the $\varphi_\mu$ dependence.
The reason is that these phase dependences are caused by the
$\tilde{t}_L$ -$\tilde{t}_R$ mixing term (eq.~(\ref{mLRterm})), where
the $\varphi_\mu$ dependence is suppressed. The $\varphi_\mu$
dependence is somewhat more pronounced for $\varphi_{A_t} \approx 0, 2\pi$
than for $\varphi_{A_t} \approx \pi$.
If the constraint $|\varphi_\mu| < 0.1$ -- $0.2$ from the EDM bounds has
to be fulfilled, then only the corresponding bands around
$\varphi_\mu = 0, \pi, 2\pi$ are allowed.
$B(b \to s\gamma)$ is in agreement with the experimental range in
almost the whole $\varphi_{A_t}$-$\varphi_\mu$ plane: only at
$\varphi_{A_t}\approx 0,2\pi$ and $\varphi_\mu\approx 0,2\pi$ it can
go up to $5\times 10^{-4}$.
In order to discuss the dependence of this branching ratio on $|A_t|$
we show in Fig.~\ref{cpneut1sps1a} (b) the contour plot of
$B(\tilde{t}_1 \to \tilde{\chi}^0_1 t)$ as a function of
$\varphi_{A_t}$ and $|A_t|$ for $\varphi_{A_b} = \varphi_\mu = 0$ and
$|A_t|=|A_b|$. Clearly, the $\varphi_{A_t}$ dependence is strongest
for large values of $|A_t|$.
The dashed lines mark the contours of $\cos\theta_{\tilde{t}}$, which
are perpendicular to the ones of $B(\tilde{t}_1 \to \tilde{\chi}^0_1 t)$
in a large domain of the parameter space.
Thus a simultaneous measurement of 
$B(\tilde{t}_1 \to \tilde{\chi}^0_1 t)$ and $\cos\theta_{\tilde{t}}$ might be 
helpful to disentangle the phase of $A_t$ from its absolute value.
As an example a measurement of 
$B(\tilde{t}_1 \to \tilde{\chi}^0_1 t)=0.6\pm 0.1$ and
$|\cos\theta_{\tilde{t}}|=0.3\pm 0.02$ would allow to determine
$|A_t|\approx 320$~GeV with an error $\Delta(|A_t|)\approx 20$~GeV and
$\varphi_{A_t}$ with a twofold ambiguity $\varphi_{A_t} \approx 0.35\pi$
or $\varphi_{A_t} \approx 1.65\pi$ with an error
$\Delta(\varphi_{A_t}) \approx 0.1\pi$.
$B(b \to s\gamma)$ is in agreement with the experimental range in
almost the whole $\varphi_{A_t}$-$|A_t|$ plane: only at
$\varphi_{A_t}\approx 0,2\pi$ and $|A_t|\gtrsim 300$~GeV it can go up to
$5\times 10^{-4}$.

In order to discuss the decays of the $\tilde{b}_1$ we choose a
scenario with a light $\tilde{b}_1$ and a light $H^\pm$ as defined in
Table~\ref{scentab}, where the $\tilde{b}_1$ production at a 800~GeV
linear collider is possible and the decay channel 
$\tilde{b}_1 \to H^- \tilde{t}_1$ is open.
We fix $\tan\beta = 30$, because for
small $\tan\beta$ the off-diagonal elements in the sbottom mixing
matrix are too small. 

\begin{figure}[ht]
\centering
\begin{picture}(16,6)
%\put(0,0){\framebox(16,6){}}

\put(-.1,0){\epsfig{file=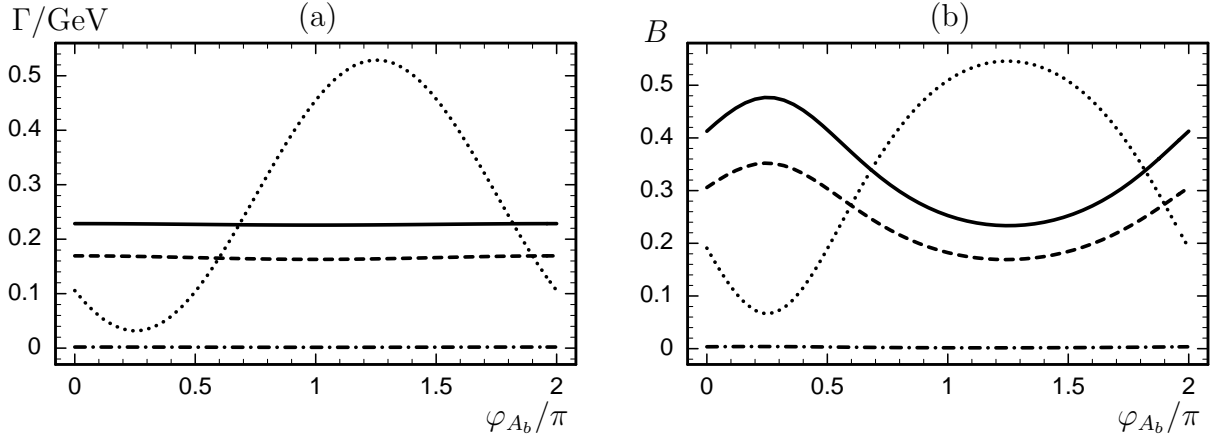}}

\end{picture}

\caption{\label{sbottomphiAb} (a) Partial decay widths and (b) branching
ratios of the decays
$\tilde{b}_1 \to \tilde{\chi}^0_1 b$ (solid),
$\tilde{b}_1 \to \tilde{\chi}^0_2 b$ (dashed),
$\tilde{b}_1 \to H^- \tilde{t}_1$ (dotted) and
$\tilde{b}_1 \to W^- \tilde{t}_1$ (dashdotted)
in the scenario with a light $\tilde{b}_1$ defined in
Table~\ref{scentab} for $|A_b|=|A_t|=600$~GeV,
$\varphi_\mu=\pi$ and $\varphi_{A_t}=\pi/4$.}

\end{figure}

We show in Fig.~\ref{sbottomphiAb} (a) and (b) the partial decay widths and
the branching ratios of $\tilde{b}_1 \to \tilde{\chi}^0_{1,2} b$,
$\tilde{b}_1 \to H^- \tilde{t}_1$ and $\tilde{b}_1 \to W^- \tilde{t}_1$
as a function of $\varphi_{A_b}$ taking $|A_b|=|A_t|=600$~GeV,
$\varphi_{\mu}=\pi$ and $\varphi_{A_t}=\pi/4$.
In the region $0.75 \pi < \varphi_{A_b} < 1.75 \pi$ the decay
$\tilde{b}_1 \to H^- \tilde{t}_1$ dominates.
The $\varphi_{A_b}$ dependence of its partial decay width is due to
that of the $\tilde{b}_R\tilde{t}_L H^-$ coupling term.
The partial decay widths of $\tilde{b}_1 \to \tilde{\chi}^0_{1,2} b$ are
essentially $\varphi_{A_b}$ independent, because the $\varphi_{A_b}$
dependence of the sbottom mixing matrix elements nearly vanishes.
The $\varphi_{A_b}$ dependence of the branching ratios
$B(\tilde{b}_1 \to \tilde{\chi}^0_{1,2} b)$ is caused by the
$\varphi_{A_b}$ dependence of the total decay width.
In the whole parameter range considered $B(b \to s\gamma)$ satisfies
the experimental limits.

\begin{figure}[ht]
\centering
\begin{picture}(16,8.5)
%\put(0,0){\framebox(16,8.5){}}

\put(-.1,0){\epsfig{file=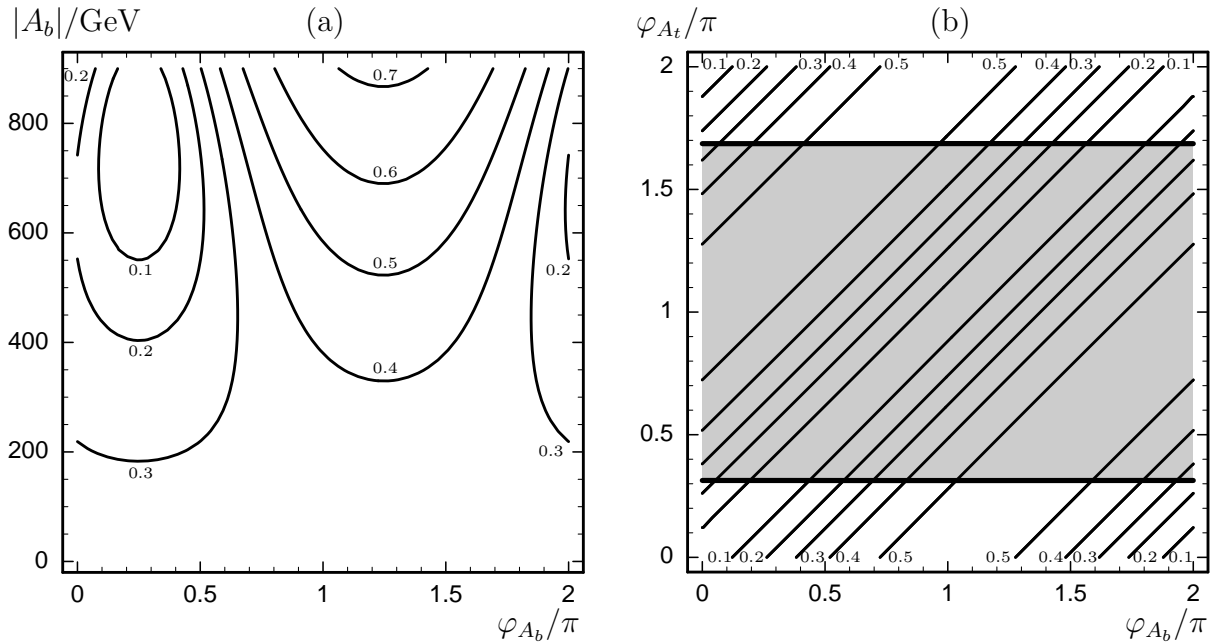}}

\end{picture}

\caption{\label{cpsbottom}Contours of $B(\tilde{b}_1 \to H^- \tilde{t}_1)$
in the scenario with a light $\tilde{b}_1$ defined in Table~\ref{scentab}
for (a)  $|A_b|=|A_t|$, $\varphi_\mu=\pi$, $\varphi_{A_t}=\pi/4$
and (b) $|A_b|=|A_t|=600$~GeV, $\varphi_\mu=\pi$.
The shaded area marks the region, which is excluded by the
experimental limit $B(b\to s \gamma) > 2.0 \times 10^{-4}$.}
\end{figure}

For large $\tan\beta$ one expects also a significant $|A_b|$
dependence of the partial decay width
$\Gamma(\tilde{b}_1 \to H^- \tilde{t}_1)$. This can be inferred from
Fig.~\ref{cpsbottom} (a), where we show the contour plot of the
branching ratio of $\tilde{b}_1 \to H^- \tilde{t}_1$ as a function of
$|A_b|$ and $\varphi_{A_b}$, taking
$|A_t| =  |A_b|$, $\varphi_{\mu}=\pi$ and $\varphi_{A_t}=\pi/4$.
The $\varphi_{A_b}$ dependence is stronger for large
values of $|A_b|$. Although Fig.~\ref{cpsbottom} (a) is similar to
Fig.~\ref{cpneut1sps1a} (b), the $|A_b|$ and $\varphi_{A_b}$ dependences in
Fig.~\ref{cpsbottom} (a) are mainly due to the phase dependence of the 
$\tilde{b}_R \tilde{t}_L H^-$ coupling.
The shifting of the symmetry axis to $\varphi_{A_b} = 1.25\pi$ is
caused by the additional phase $\varphi_{A_t}=\pi/4$.
Contrary to the mixing in the stop sector $\cos\theta_{\tilde{b}}$ is
nearly independent of $|A_b|$ and $\varphi_{A_b}$. Therefore the knowledge
of $\cos\theta_{\tilde{b}}$ does not help to disentangle the phase of
$A_b$ from its absolute value.

In Fig.~\ref{cpsbottom} (b) we show the contour lines of
$B(\tilde{b}_1 \to H^- \tilde{t}_1)$ as a function of $\varphi_{A_b}$
and $\varphi_{A_t}$, for $|A_t| =  |A_b| = 600$~GeV, $\varphi_{\mu}=\pi$
and the other
parameters (except $\varphi_{A_t}$) as in Fig.~\ref{cpsbottom} (a).
As can be seen, the $\varphi_{A_b}$-$\varphi_{A_t}$
correlation is quite strong.
The shaded area marks the region which is experimentally excluded
because of $B(b\to s \gamma) < 2.0\times 10^{-4}$.
Note, that the constraints from $B(b\to s \gamma)$ are only fulfilled
for a small range of values of $\varphi_{A_t}$
in the given scenario with $m_{H^\pm}=150$~GeV.

\section{Conclusion}

We have shown that the effect of the CP-violating phases of the
supersymmetric parameters $A_t$, $A_b$ and $\mu$ on CP-conserving
observables such as the branching ratios of $\tilde{t}_1$ and
$\tilde{b}_1$ decays can be strong in a large region of the
MSSM parameter space.
Especially the branching ratios of the $\tilde{t}_1$ can show a pronounced
dependence on $\varphi_{A_t}$ in all decay channels.
The dependence of the partial decay widths of the $\tilde{b}_1$ on
$\varphi_{A_b}$ are in general quite small with exception of decays into
final states containing Higgs bosons. Nevertheless the resulting
branching ratios show a clear phase dependence.
This could have an important impact on the search for $\tilde{t}_1$
and $\tilde{b}_1$ at a future $e^+e^-$ linear collider and on the
determination of the MSSM parameters, especially of $A_t$ and $A_b$
which are not easily accessible otherwise.

\section*{Acknowledgements}

This work is supported by the `Fonds zur F\"orderung der
wissenschaftlichen For\-schung' of Austria, FWF Projects No.~P13139-PHY
and No.~P16592-N02
and by the European Community's Human Potential Programme
under contract HPRN-CT-2000-00149.
W.P.\ has been supported by the Erwin Schr\"odinger fellowship No.~J2272 of
the `Fonds zur F\"orderung der wissenschaftlichen Forschung' of Austria
and partly by the Swiss `Nationalfonds'.

\end{document}